\documentclass[fleqn,12pt,twoside]{article}
\usepackage{espcrc1}
\usepackage{graphicx}

\newcommand{\dd}{{\rm d}}

\title{Multiplicity distributions for jet parton showers in a medium}

\author{Nicolas Borghini\address[CERN]{Physics Department, Theory Division, 
    CERN, CH-1211 Geneva 23, Switzerland}
  and 
  Urs Achim Wiedemann\addressmark[CERN]$^,$%
    \address{Physics Department, University of Bielefeld, D-33501 Bielefeld, 
      Germany}}

\runtitle{Multiplicity distributions for jets in a medium}
\runauthor{N. Borghini}
 
\begin{document}

\maketitle

\begin{abstract}
The ``jet-quenching'' interpretation of suppressed high-$p_T$ hadron production 
at RHIC implies that jet multiplicity distributions and jet-like particle 
correlations in heavy-ion collisions at RHIC and LHC differ strongly from those 
seen at $e^+e^-$ or $pp$ ($p\bar p$) colliders. 
Here, we present an approach for describing the changes induced by the medium, 
which implements jet quenching as a probabilistic medium-modified parton shower,
treating leading and subleading parton splittings on an equal footing. 
We show that the strong suppression of single inclusive hadron spectra measured 
in Au--Au collisions at RHIC implies a characteristic distortion of the single 
inclusive distribution of soft partons inside the jet. 
We determine, as a function of jet energy, to what extent the soft jet fragments
can be measured above some momentum cut. 
\end{abstract}

\section{INTRODUCTION}

One of the major discoveries of the first years of nucleus--nucleus collisions 
at RHIC is the observed deficit in the production of high transverse momentum 
particles if compared to $pp$ collisions~\cite{Jacobs:2004qv}. 
The measurements strongly support the ``jet quenching'' interpretation, that
high-$p_T$ partons traveling through the dense medium 
created in the collisions lose a significant fraction of their energy, through 
an enhanced radiation of gluons induced by the medium~\cite{Baier:2000mf,%
  Gyulassy:2003mc,Kovner:2003zj}, before they hadronize outside the medium. 
Models based on this gluon radiation thus successfully account for the measured 
inclusive single-particle spectra and leading back-to-back two-particle 
correlations~\cite{Dainese:2004te,Eskola:2004cr}. 

Despite these successes of models based on medium-induced radiative energy loss 
--- which assume either multiple soft-momentum transfers~\cite{Baier:1996sk,%
  Salgado:2003gb} or single hard scattering~\cite{Gyulassy:2003mc} from the 
medium --- there is much room for technical improvements.
In particular, standard formulations of parton energy loss consider only the 
medium-induced splitting of the leading parton, thus treating the splitting of 
leading and subleading partons differently. 
This makes it questionable to apply these approaches to the many measurements of
jets and jet-like characteristics, such as particle-triggered multiplicity 
distributions, that are now gradually becoming available. 
Also, the current state of the art in applying medium-modified parton splittings
involves numerically significant ad hoc corrections to maintain energy-momentum 
conservation.

In Ref.~\cite{Borghini:2005em}, we have proposed an alternative formulation of 
medium-induced parton energy loss. Our approach is the first that treats both 
leading and subleading parton splittings on an equal footing. 
In addition, it conserves energy-momentum at each splitting. 
One of the main motivations for this approach is the possibility that these 
features may be more important than the precise treatment of coherence effects 
in medium-induced gluon emission, in particular for the discussion of high-$p_T$
particle correlations (for which an interesting, alternative extension of parton
energy loss was proposed in Refs.~\cite{Majumder:2004wh,Majumder:2004pt}) and 
multiplicity distributions inside jets.

\section{FORMALISM}
\label{s:formalism}

Jets in $e^+e^-$ or $pp$/$p\bar p$ collisions provide stringent tests of QCD.
Color coherence, which results in destructive interference between partons with 
small momentum fractions $x=p/E_{\rm jet}$ inside a shower, leads to unique 
predictions for the single inclusive distribution $D(x,Q^2)$ of partons inside a
jet of energy and virtuality $E_{\rm jet} \sim Q$~\cite{Mueller:1982cq,%
  Dokshitzer:1988bq}.
The interference is actually equivalent, to double and single logarithmic 
accuracy in $\ln(1/x)$ and $\ln(Q/\Lambda_{\rm eff})$ where $\Lambda_{\rm eff}$
is an infrared cutoff, to an angular-ordered probabilistic parton shower with 
leading-order splitting functions. 
The resummed Modified Leading Logarithmic Approximation (MLLA) describes, to 
next-to-leading order $\sqrt{\alpha_S}$, the measured longitudinal distributions
of hadrons $D^h(x,Q^2)$ in collisions over a wide energy range~\cite{%
  Braunschweig:1990yd,Abbiendi:2002mj}, provided each parton is mapped locally 
onto a hadron (``Local Parton--Hadron Duality'', LPHD) with a proportionality 
factor $\sim {\cal O}(1)$.
As an example, Fig.~\ref{fig:limitspectra} shows the distribution $D^h(x,Q^2)$ 
inside 17.5~GeV jets, a typical jet energy at RHIC, measured in $e^+e^-$ 
collisions by the TASSO Collaboration~\cite{Braunschweig:1990yd}, together with 
the MLLA prediction.

We have developed a formalism for medium-distorted jets which reduces to the
MLLA baseline in the absence of a medium~\cite{Borghini:2005em}.
With respect to the standard formalism of parton energy loss used currently in 
the analysis of RHIC data, the approach involves different approximations.
Present model comparisons to RHIC data start with a medium-modified energy 
distribution of gluons $\dd I^{\rm tot} = \dd I^{\rm vac} + \dd I^{\rm med}$. 
Here, the vacuum contribution is double logarithmic $\dd I^{\rm vac} = 
  \frac{\alpha_s}{\pi^2} \frac{\dd\omega}{\omega} \frac{\dd{\bf k}}{{\bf k}^2}$,
and its integral over ${\bf k}$ gives rise to the leading $\ln Q^2$ term in the 
DGLAP evolution equation. 
In contrast, the ${\bf k}$-integration of $\dd I^{\rm med}$ is infrared- and 
ultraviolet-safe~\cite{Salgado:2003gb}, and leads to a geometrically enhanced 
($\sim L$) ``higher twist'' contribution $\propto\hat{q}L/Q^2$. 
However, these qualitatively different $Q^2$-dependences do not enter current 
comparisons to RHIC data. 
Instead, one considers the ${\bf k}$-integrated spectrum 
$\omega \frac{\dd I^{\rm med}}{\dd\omega}$ as an additional source of gluon 
radiation, whose dependence on $Q^2$ is neglected. 
Present approximations also do not include the further medium-induced splitting
processes of subleading partons in the shower. 

One improvement of this state of the art may be to replace the double 
differential $\dd I^{\rm vac}$ by $\dd I^{\rm tot}$ in {\em all\/} leading and 
subleading splittings of a medium-modified parton shower. 
This requires a Monte-Carlo approach, which we intend to develop in future work.
For the first analytical study, reported here, we have employed an additional
approximation: We replaced the ${\bf k}$-integrated medium distribution 
$\omega \frac{\dd I^{\rm med}}{\dd\omega}$ by a constant $f_{\rm med}$. 
In the kinematic regime tested at RHIC, this turns out to be a fair 
approximation.
We have then used $\omega \frac{\dd I^{\rm med}}{\dd\omega}$ on the same level 
as $\omega \frac{\dd I^{\rm vac}}{\dd\omega}$, i.e., as a leading logarithmic 
correction~\cite{Borghini:2005em}.
With this ansatz, our formalism ensures energy-momentum conservation at each 
parton splitting, and treats all leading and subleading parton splittings on the
same footing. 
The price we pay for an analytical treatment is the 
inaccurate handling of the $Q^2$-dependence (which is not dealt with more 
properly in other formalisms) and the approximate treatment of the 
$\omega$-dependence. 
In a Monte-Carlo formulation of the same problem, we expect to remove 
the latter deficiencies. 
Moreover, the analytical results presented here will serve as a powerful 
consistency check for the correct implementation of splitting processes in the 
Monte-Carlo simulation.

\section{PHENOMENOLOGICAL PREDICTIONS}
\label{s:results}

In our analytical approach, we approximate the medium-induced contribution 
$\omega \frac{\dd I^{\rm med}}{\dd\omega}$ by a constant $f_{\rm med}$ in the 
kinematically relevant range of $\omega$. 
Medium-modified parton splitting functions then differ from the standard ones 
by multiplying their singular contributions with $(1 + f_{\rm med})$. 
This formulation allows us to follow the same line of technical arguments, used 
for the calculation of jet multiplicity distributions in the absence of a 
medium~\cite{Dokshitzer:1988bq}.
\begin{figure}
  \begin{minipage}{0.49\linewidth}
    \centerline{\includegraphics*[width=\linewidth]{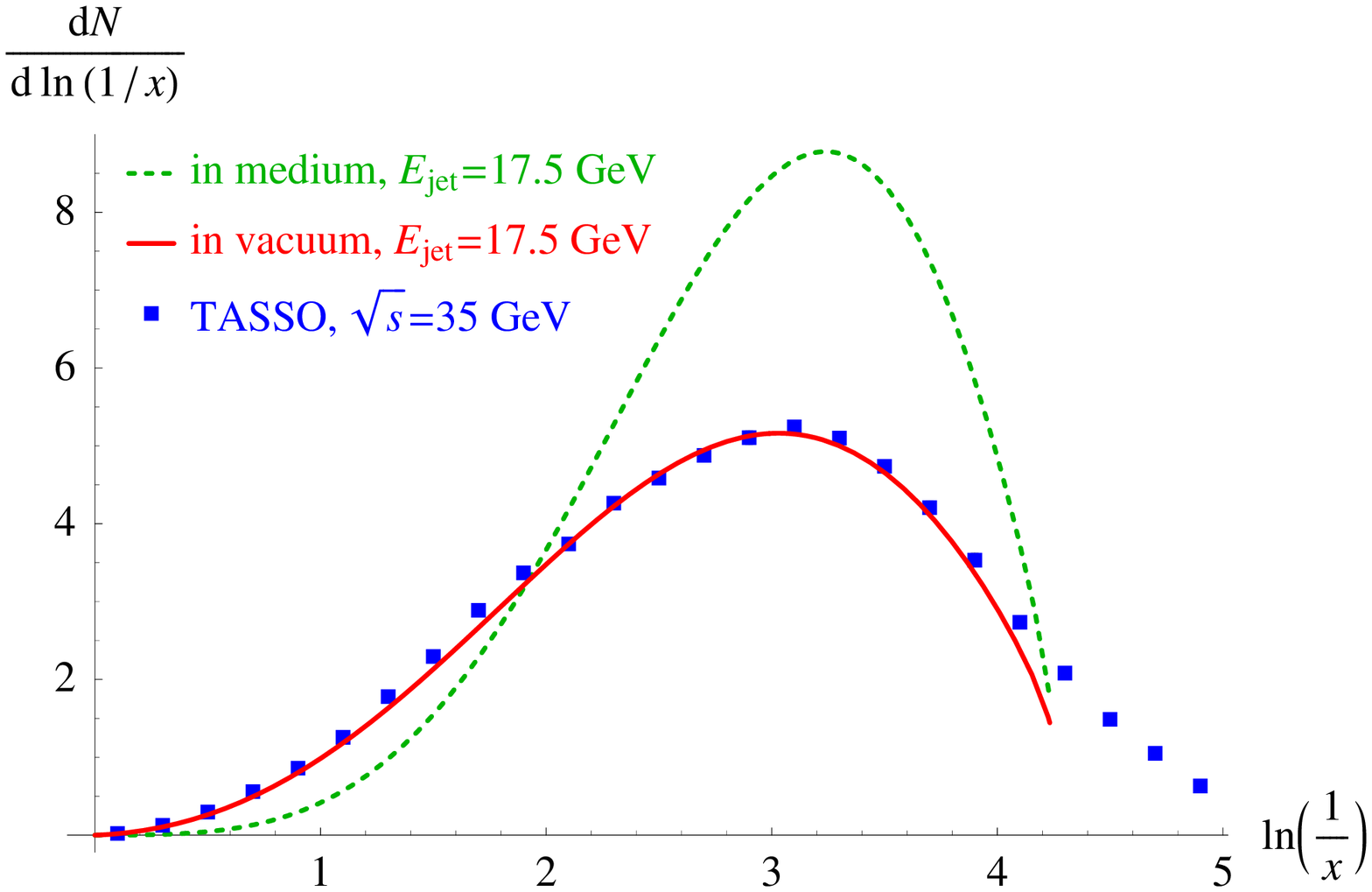}}
    \vspace{-5mm}
    \caption{Single inclusive distribution of hadrons vs. $\ln\,(E_{\rm jet}/p)$
      for $E_{\rm jet}=17.5$~GeV jets measured by TASSO in $e^+e^-$ collisions 
      and within MLLA (solid curve: $f_{\rm med}=0$, dashed curve: 
      $f_{\rm med}=0.8$).\label{fig:limitspectra}}
  \end{minipage}\hfill
  \begin{minipage}{0.49\linewidth}
    \centerline{\includegraphics*[width=\linewidth]{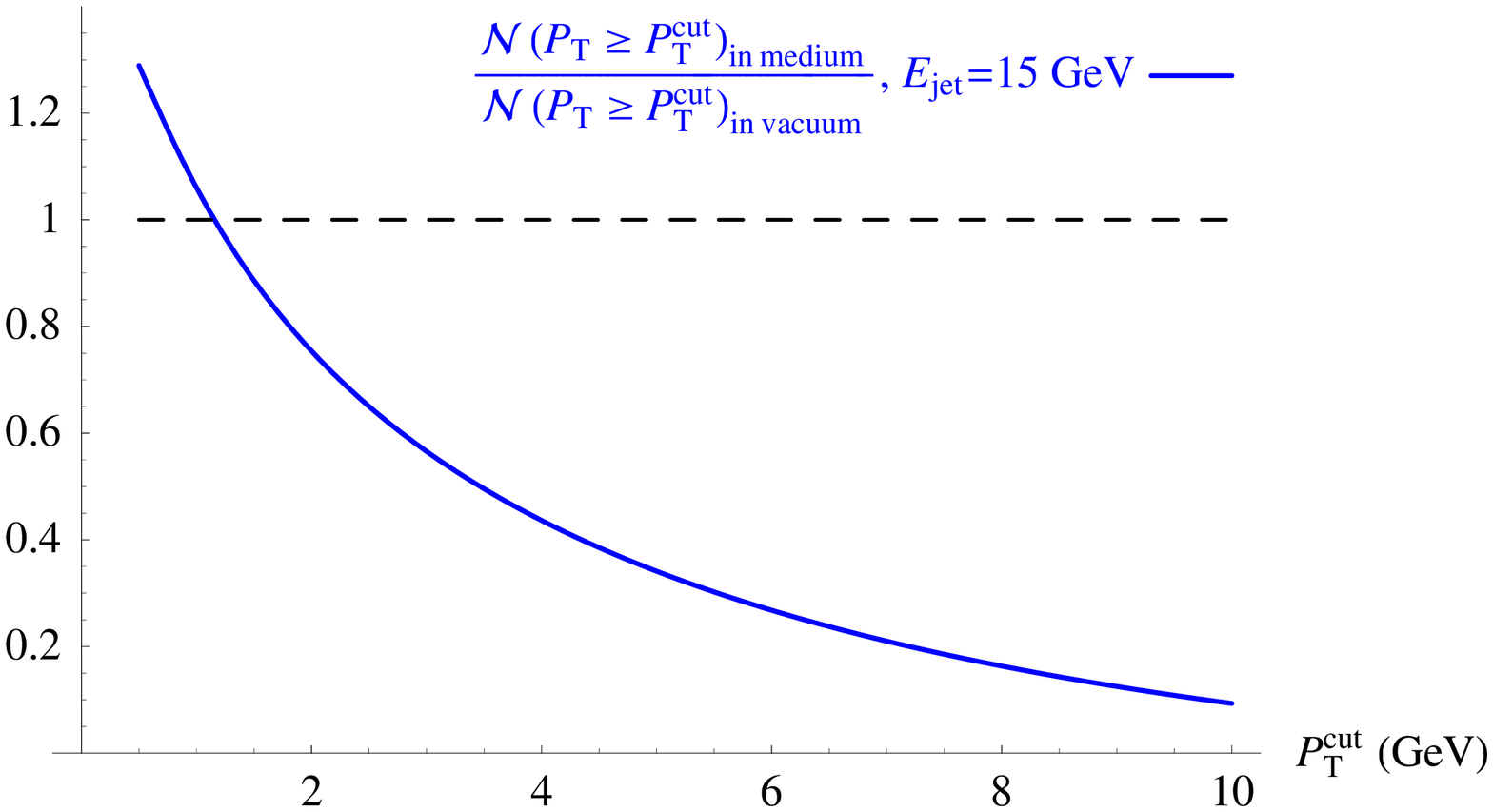}}
    \vspace{-8.5mm}
    \caption{Ratio of the multiplicities inside a jet in medium and in vacuum 
      vs. the soft momentum cut above which the multiplicity is measured.
      \label{fig:assocmult}}
  \end{minipage}%
  \vspace{-5mm}
\end{figure}
In Fig.~\ref{fig:limitspectra}, one sees the resulting in-vacuum and 
medium-distorted distributions $D^h(x,Q^2)$, plotted as a function of 
$\ln(1/x)$, for jets with $E_{\rm jet}=17.5$~GeV. We chose the value
$f_{\rm med}=0.8$ to obtain in the same formalism a fair description 
of the suppression of high-$p_T$ hadrons~\cite{Borghini:2005em}. 
Figure~\ref{fig:limitspectra} illustrates the effect of the medium, which 
results in a depletion of the number of particles at large $x$, and 
correspondingly a largely enhanced emission of particles at small $x$: 
due to energy-momentum conservation in the parton 
shower, the energy which in the vacuum is taken by a single large-$x$ parton is 
redistributed over many small-$x$ partons in the presence of a medium. 

Given the longitudinal multiplicity distribution inside a jet, a straightforward
integration yields the number of particles inside the jet with momenta larger 
than a given cut, ${\cal N}(P_T\geq P_T^{\rm cut})$. 
We can then compute this multiplicity for jets with the same given energy both 
in the presence of medium effects (in which case, the cut gives some control on
the high-multiplicity soft background of heavy-ion collisions at RHIC or LHC 
over which the jet develops) and in vacuum, and form their ratio.
The latter, shown in Fig.~\ref{fig:assocmult} for jets with $E_{\rm jet}=15$~GeV
and medium effects modeled by $f_{\rm med}=0.8$, is seen to be smaller than 1 
for $P_T^{\rm cut}>1.5$~GeV, while the medium-induced enhancement in soft 
particles becomes dominant for smaller values of the momentum cut.
Although the medium-enhanced part of the multiplicity is {\em a priori\/} buried
in the soft background of Au--Au collisions at RHIC, this crossover value lies 
close to that reported by the STAR Collaboration in attempts at measuring the 
excess~\cite{Adams:2005ph}. 
Note, however, that in our calculation the jet energy is known and fixed, while 
experimentally it spans a rather wide range.

\section{CONCLUSION}

We have reported a first step towards a description of parton showers in a 
medium, which conserves energy-momentum at each parton splitting, and deals in a
symmetric way with all partons in the shower~\cite{Borghini:2005em}. 
Our simplified analytical formalism is able to reproduce at least 
semi-quantitatively several characteristic features of RHIC data, such as the 
leading-hadron suppression and the enhanced soft-particle distribution 
associated to high-$p_T$ trigger particles.
In the future, a Monte-Carlo implementation of this approach may be of great 
interest, since it would allow one to include more realistic medium-induced 
splitting functions. 
Also, it would allow us to address the issue whether and how medium-induced 
parton energy loss depends on virtuality. 
This question is of particular importance for addressing the logarithmically 
wide kinematic $p_T$ range accessible at the LHC.
\medskip \medskip

We thank Carlos Salgado for stimulating critical discussions.

\end{document}